\documentclass[aps,prd,showpacs,twocolumn,groupedaddress]{revtex4}
\usepackage{graphicx}

\begin{document}

\title{Quark motional effects on the inter-quark potential in baryons}

\author{Arata~Yamamoto}
\affiliation{Faculty of Science, Kyoto University, Kitashirakawa, Sakyo, Kyoto 606-8502, Japan}

\author{Hideo~Suganuma}
\affiliation{Faculty of Science, Kyoto University, Kitashirakawa, Sakyo, Kyoto 606-8502, Japan}

\date{\today}

\begin{abstract}
We study the heavy-heavy-light quark ($QQq$) system in a non-relativistic potential model, and investigate the quark motional effect on the inter-two-quark potential in baryons. 
We adopt the Hamiltonian with the static three-quark potential which is obtained by the first-principle calculation of lattice QCD, rather than the two-body force in ordinary quark models.
Using the renormalization-group inspired variational method in discretized space, we calculate the ground-state energy of $QQq$ systems and the light-quark spatial distribution.
We find that the effective string tension between the two heavy quarks is reduced compared to the static three-quark case.
This reduction of the effective string tension originates from the geometrical difference between the inter-quark distance and the flux-tube length, and is conjectured to be a general property for baryons.
\end{abstract}

\pacs{12.39.Jh, %quark model
12.39.Pn,       %potential model
14.20.Lq, 14.20.Mr}

\maketitle

\section{Introduction}

In hadron physics, the inter-quark interaction is one of essential properties reflecting the non-perturbative gluon dynamics based on the SU(3) gauge symmetry.
Particularly the quark confinement is a nontrivial, non-perturbative, and unsolved problem even in static quark cases.
The interaction between finite-mass quarks would include more complicated effects.
Additionally, in contrast to mesons, quarks in baryons have various kinds of the motion, configuration, and so on, and such a quark degree of freedom would be important for the inter-quark interaction.
The inter-two-quark potential in baryons can be influenced by such nontrivial effects of the other quark.

To investigate such finite-mass-quark ``motional" effects on the effective inter-two-quark potential, we study the heavy-heavy-light quark ($QQq$) system in this work.
We treat the two heavy quarks in  the $QQq$ system as infinitely heavy, keeping the light-quark mass finite.
We artificially change the inter-heavy-quark distance $R$, and calculate the energy of the $QQq$ system as a function of $R$, which we call the $QQq$ potential.
This $QQq$ potential includes not only the gluonic effect but also the nontrivial light-quark effect, and would differ from the static three-quark ($3Q$) potential.

From our previous lattice QCD study \cite{YaUP}, the $QQq$ potential is almost described as a linear and one-gluon-exchange Coulomb potential, like the quark-antiquark ($Q\bar{Q}$) potential.
However, the effective string tension between two heavy quarks is reduced compared to the static $Q\bar{Q}$ or $3Q$ case, 0.89 GeV/fm.
This reduction would be the result from nontrivial light-quark effects.
Lattice QCD is a powerful tool for calculating hadron masses, potentials, and several QCD properties, but has a difficulty to clarify the light-quark wave function. 
Instead, in this paper, we employ a simple potential model, and calculate the $QQq$ potential and the spatial distribution of the light-quark wave function, in order to explain the reduction mechanism of the effective string tension.

In most of ordinary quark models and potential models, the quark confining force in baryons is the simple two-body force \cite{Ca86,Ok80,Vi04}.
However, it is recently found that the confinement potential in baryons is expressed by the Y-type flux tube picture \cite{Ta0102,Ok05,Ic03,Fa91,Br95,Co04}.
For more accurate calculations, quark models and potential models need to treat the confinement force in baryons as the three-body force.
In particular, since we are now interested in the light-quark effects on the inter-quark potential, the accurate treatment for the three-quark interaction is important.
Hence, we adopt the three-quark potential from the lattice QCD result, which is the non-perturbative first-principle calculation.

As an example of realistic $QQq$ baryons, the doubly charmed baryon is experimentally observed.
In 2002, the SELEX Collaboration at Fermilab observed $\Xi _{cc}^+(dcc)$ through a weak decay $\Xi _{cc}^+ \to \Lambda _c^+ K^- \pi ^+$  \cite{Ma02}.
They also confirmed another decay process $\Xi _{cc}^+ \to p D^+ K^- $ \cite{Oc05}. 
In their experiments, the mass is measured about 3519 MeV.
Doubly charmed baryons are also theoretically investigated in lattice QCD \cite{Le0102}, potential model \cite{Ru75,Vi04}, and other approaches \cite{Br05}.

We already know analogous systems to this $QQq$ baryon, for example the H$_2^+$ ion in molecular physics, which is also the bound state of two heavy and one light particles.
However, the interaction in the ground-state $QQq$ system is attractive, while the H$_2^+$ ion includes Coulomb repulsive force between the two protons.
Additionally, as stated above, the confining force in the $QQq$ baryon is a purely three-body force, not a sum of two-body forces, and has the characteristic geometrical structure.
Then the $QQq$ potential is expected to include non-trivial effects from this feature. 

In this paper, we study finite-mass-quark effects on the inter-two-quark potential in baryons through the investigation of $QQq$ systems with a non-relativistic potential model.
The article is organized as follows.
In Sec.~II, we introduce the Hamiltonian with the confinement potential obtained from recent lattice QCD calculations, and present the formalism of the renormalization-group inspired variational calculation in a discretized space to solve the Schr$\ddot{{\rm o}}$dinger equation.
In Sec.~III, we show the numerical results of the $QQq$ potential, and discuss the finite-mass-quark effect for the reduction of the effective string tension between the two heavy quarks in $QQq$ systems.
Sec.~IV is devoted to the summary and the conclusion.

\section{Formalism}
\subsection{Hamiltonian}

The Hamiltonian of three quarks in baryons is constructed from kinetic terms and a three-quark interaction term as
\begin{eqnarray}
H=\sum_{i=1}^3 T_i+V(\vec{r}_1,\vec{r}_2,\vec{r}_3).
\end{eqnarray}
In the $QQq$ system, we treat two heavy quarks as infinitely heavy particles, and one light quark as a non-relativistic constituent quark with the constituent mass $M_q$.
The non-relativistic quark model is one of the successful models to describe baryons even for the light-quark sector, and have been used for the study of baryons by many theoretical physicists even at present \cite{Ya06}.
Apart from irrelevant constants, the Hamiltonian is simplified as
\begin{eqnarray}
\label{Hami}
H=M_q-\frac{1}{2M_q}\frac{\partial^2}{\partial \vec{r}_3^2}+V(\vec{r}_1,\vec{r}_2,\vec{r}_3),
\end{eqnarray}
where the subscripts 1, 2, and 3 mean the two heavy quarks and the light quark, respectively.
Although we adopt the non-relativistic formalism, we still call this finite-mass quark the ``light" quark in this paper.

As the three-quark interaction $V(\vec{r}_1,\vec{r}_2,\vec{r}_3)$, to treat the light-quark effect as precisely as possible, we adopt the lattice QCD result of the static $3Q$ potential \cite{Ta0102}.
This $3Q$ potential includes the confining potential as the three-body force, instead of the simple sum of the two-body force in ordinary quark models.
The static $3Q$ potential obtained by quenched lattice QCD is
\begin{eqnarray}
\label{V123}
V(\vec{r}_1,\vec{r}_2,\vec{r}_3)=\sigma _{3Q}L_{\rm min}-\sum _{i< j}\frac{A_{3Q}}{r_{ij}}+C_{3Q},\\
\sigma _{3Q} \simeq 0.89 \ {\rm GeV/fm}, \quad A_{3Q} \simeq 0.13,
\label{SA3Q}
\end{eqnarray}
where $r_{ij}=|\vec{r}_i-\vec{r}_j|$, and these values in Eq.(\ref{SA3Q}) are related to the $Q\bar{Q}$ case as $\sigma _{3Q} \simeq \sigma_{Q\bar{Q}}$ and $A_{3Q} \simeq A_{Q\bar{Q}}/2$ \cite{Ta0102}.
Since we are not interested in a constant shift of the energy in the non-relativistic formalism, we set $C_{3Q}=0$. 
The symbol $L_{\rm min}$ is the length of the color flux tube minimally connecting the three quarks, which is described as follows.
When all the angles of the $3Q$ triangle is less than $2\pi /3$, the Y-type flux tube is formed and
\begin{eqnarray}
L_{\rm min}&=&\frac{1}{\sqrt{2}}\Big[ r_{12}^2+r_{13}^2+r_{23}^2\nonumber\\
&+&\sqrt{3(r_{12}+r_{13}+r_{23})(-r_{12}+r_{13}+r_{23})}\nonumber\\
&\times& \sqrt{(r_{12}-r_{13}+r_{23})(r_{12}+r_{13}-r_{23})}\Big]^{1/2}.
\end{eqnarray}
When one of the angles of the $3Q$ triangle exceeds $2\pi /3$,
\begin{eqnarray}
L_{\rm min}=r_{12}+r_{13}+r_{23}-{\rm max}(r_{12},r_{13},r_{23}).
\end{eqnarray}
Once the heavy-quark coordinates $\vec{r}_1$ and $\vec{r}_2$ are fixed, the interaction depends only on the light-quark coordinate $\vec{r}_3$.
We can calculate the ground-state light-quark wave function $\psi (\vec{r}_3)$ with the variational principle of the energy of the system
\begin{eqnarray}
E(R)=\frac{\int d^3r_3 \psi^*(\vec{r}_3) H \psi(\vec{r}_3)}{\int d^3r_3 |\psi (\vec{r}_3)|^2}.
\end{eqnarray}
We determine the ground-state $QQq$ potential $V_{QQq}(R)$ by minimizing $E(R)$.

Here, we comment on the three-quark interaction $V(\vec{r}_1,\vec{r}_2,\vec{r}_3)$ with finite-mass quarks.
The finite-mass effect changes the simple Coulomb interaction to the Fermi-Breit interaction \cite{Ru75}, which includes spin-spin and spin-orbit interactions.
These relativistic corrections are suppressed by the inverse of the quark mass.
In the $QQq$ system with infinitely heavy quarks, most of them give zero contributions as $1/M_Q\to 0$, and the interaction remains the simple form.
However, the finite-mass effect on the quark confinement potential is unknown.
Here, for simplicity, we assume that $V(\vec{r}_1,\vec{r}_2,\vec{r}_3)$ in the $QQq$ system can be written with the static one.
Equivalently, the finite-mass effect is assumed to be taken only via the light-quark wave function spreading.

\subsection{The renormalization-group (RG) inspired variational calculation}

\begin{figure}[t]
\begin{center}
\includegraphics[scale=0.4]{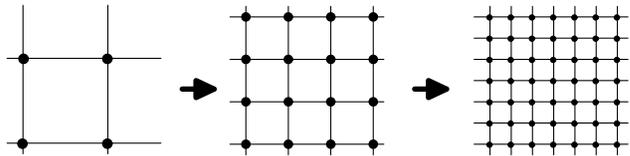}
\caption{\label{fig1}
The schematic figure of the renormalization group (RG) inspired variational calculation.
The finer-mesh calculation is done with the initial condition constructed from the rougher-mesh result.
}
\end{center}
\end{figure}

We exactly solve the energy variational problem in discretized space.
We take a cylindrical coordinate $(\rho ,\theta ,z)$, and locate the two heavy quarks on $(0,0,R/2)$ and $(0,0,-R/2)$.
The ground-state light-quark wave function is mirror symmetric to the $z=0$ plane and rotational symmetric around the $z$-axis.
Thus, we have only to calculate on the two-dimensional plane $(\rho ,z)$ ($\rho \ge 0$, $z\ge 0$).
We discretize the space with an ``isotropic" mesh as $\Delta \rho =\Delta z$, and vary the light-quark wave function on each sites to minimize $E(R)$.
This is equivalent to solving the Schr$\ddot{{\rm o}}$dinger equation exactly, and we have no Ansatz about the functional form of the light-quark wave function.

\begin{figure*}[t]
\includegraphics[scale=1.4]{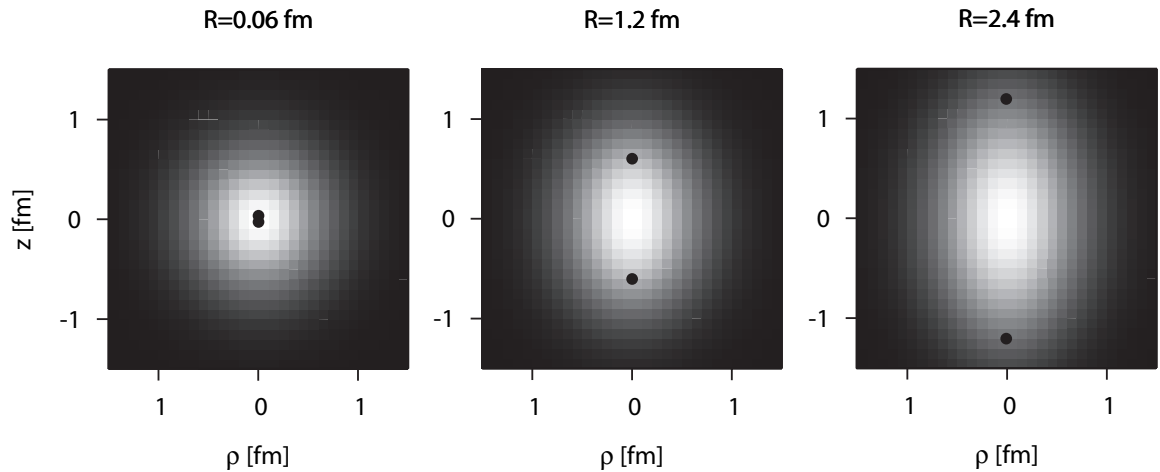}
\caption{\label{fig2}
The light-quark spatial distribution $|\psi (\vec{r}_3)|^2$ for $M_q=330$ MeV with $R=0.06$ fm (left), $R=1.2$ fm (center), and $R=2.4$ fm (right).
The brighter region has higher probability, and the black circles denote the positions of the heavy quarks.
The figure is a part of the whole volume, and the actual calculation is performed in enough large volume.
}
\end{figure*}

\begin{table*}[t]
\caption{\label{tab1}
Numerical results for the different light-quark masses $M_q$ and the different mesh sizes $\Delta z(=\Delta \rho)$.
The calculation listed here is done with the 128$\times$128-mesh.
The omitted length unit is fm.
}
\begin{ruledtabular}
\begin{tabular}{ccccccccc}
$M_q$ [GeV]& $\Delta z$ 
& $\sqrt{\langle x^2\rangle}$($R$=0.06) & $\sqrt{\langle x^2\rangle}$($R$=1.2)
& $\sqrt{\langle z^2\rangle}$($R$=0.06) & $\sqrt{\langle z^2\rangle}$($R$=1.2)
& $\langle L_{\rm min}\rangle$($R$=0.06) & $\langle L_{\rm min}\rangle$($R$=1.2)\\
\hline
0.33& 0.08 & 0.41 & 0.42 & 0.43 & 0.53 & 0.67 & 1.61 \\
	& 0.10 & 0.42 & 0.41 & 0.45 & 0.53 & 0.67 & 1.60 \\
	& 0.12 & 0.40 & 0.40 & 0.46 & 0.55 & 0.68 & 1.60 \\
0.50& 0.08 & 0.36 & 0.36 & 0.39 & 0.47 & 0.58 & 1.53 \\
    & 0.10 & 0.35 & 0.35 & 0.39 & 0.49 & 0.59 & 1.53 \\
1.0 & 0.05 & 0.27 & 0.29 & 0.30 & 0.42 & 0.46 & 1.44 \\
	& 0.10 & 0.27 & 0.28 & 0.32 & 0.42 & 0.47 & 1.43 \\
2.0 & 0.05 & 0.20 & 0.23 & 0.23 & 0.37 & 0.35 & 1.36 \\
    & 0.08 & 0.20 & 0.22 & 0.24 & 0.37 & 0.36 & 1.36 \\
\end{tabular}
\end{ruledtabular}
\end{table*}

We adopt the variational calculation inspired by the renormalization group (RG) method.
The schematic procedure is shown in Fig.~\ref{fig1}, and its concrete process is as follows.
First, we start with a 2$\times$2 mesh and the spacings $\Delta \rho^{(1)}$ and $\Delta z^{(1)}$.
We minimize $E(R)$, and then obtain the 2$\times$2-solution $\psi^{(1)} (l,m)$.
Here the site $(l,m)$ corresponds to $(\rho,z)=(l\Delta\rho^{(1)},m\Delta z^{(1)})$ ($l,m \in {\bf Z}$).
Next, we turn to a 4$\times$4 mesh with the twice finer mesh size, starting from the initial condition $\psi^{(2)}_0 (l',m')$ from the 2$\times$2-mesh solution.
The $2^{n+1}\times 2^{n+1}$-mesh initial condition is set from the $2^n\times 2^n$-mesh solution, as
\begin{eqnarray}
&&\Delta\rho^{(n+1)} = \frac{1}{2}\Delta\rho^{(n)},\\
&&\Delta z^{(n+1)} = \frac{1}{2}\Delta z^{(n)},\\
&&\psi^{(n+1)}_0 (2l,2m)= \psi^{(n)} (l,m),\\
&&\psi^{(n+1)}_0 (2l-1,2m) \nonumber\\
&&\qquad = \frac{1}{2}\{ \psi^{(n)} (l,m)+\psi^{(n)} (l-1,m)\},\\ 
&&\psi^{(n+1)}_0 (2l,2m-1)\nonumber\\
&&\qquad  = \frac{1}{2}\{ \psi^{(n)} (l,m)+\psi^{(n)} (l,m-1)\},\\
&&\psi^{(n+1)}_0 (2l-1,2m-1) \nonumber\\
&&\qquad = \frac{1}{2}\{ \psi^{(n)} (l,m)+\psi^{(n)} (l-1,m-1)\}.
\end{eqnarray}
We repeat this procedure $N$ times, finally obtain the $2^N\times 2^N$-mesh solution with the spacings $\Delta\rho\equiv\Delta\rho^{(N)}$ and $\Delta z\equiv\Delta z^{(N)}$.
With this RG inspired variational calculation, the solution is expected to converge rapidly to the absolute minimum.
To estimate the discretization error and the finite-volume effect, we calculate with several mesh sizes and mesh numbers.

\section{Results}
\subsection{The $QQq$ potential}

Three examples of the obtained light-quark spatial distribution are shown in Fig.~\ref{fig2}.
When $R$ becomes large, the light-quark distribution is broadened in the $z$ direction.
To estimate the size of the light-quark spreading, we calculate $\sqrt{\langle z^2\rangle}$ and $\sqrt{\langle x^2\rangle}=\sqrt{\langle \rho^2\rangle /2}$ from the obtained wave function.
Some typical examples of these values are shown in Table \ref{tab1}.
$\sqrt{\langle x^2\rangle}$ is almost $R$-independent, and is about 0.4 fm for $M_q$=330 MeV.

\begin{figure}[t]
\begin{center}
\includegraphics[scale=1.2]{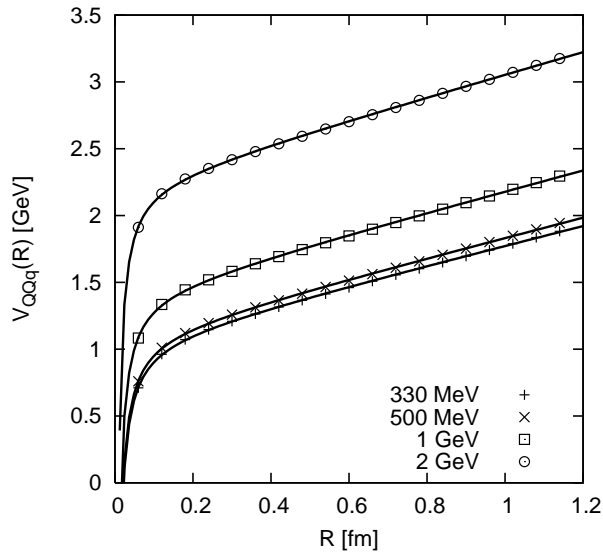}
\caption{\label{fig3}
The $QQq$ potential $V_{QQq}(R)$ from the variational calculation.
The different four symbols denote the result with the different four light-quark masses $M_q=0.33$, 0.50, 1.0, 2.0 GeV.
The solid curves are the best-fit functions of Eq.~(\ref{VQQq}).
}
\end{center}
\end{figure}

\begin{table}
\caption{\label{tab2}
The best-fit parameters $\sigma_{\rm eff}$, $A_{\rm eff}$, and $C_{\rm eff}$ in Eq.~(\ref{VQQq}) with the fit range of $R\le 1.2$ fm.
The 128$\times$128-mesh is used.
This $C_{\rm eff}$ differs from $C_{\rm eff}$ of the lattice result in Ref.~\cite{YaUP}.
}
\begin{ruledtabular}
\begin{tabular}{cccccc}
$M_q$ [GeV]& $\Delta z$ [fm] & $\sigma_{\rm eff}$ [GeV/fm] & $A_{\rm eff}$ & $C_{\rm eff}$ [GeV]\\
\hline
0.33& 0.08 & 0.74 & 0.12 & 1.06\\
	& 0.10 & 0.73 & 0.12 & 1.07\\
	& 0.12 & 0.73 & 0.12 & 1.07\\
0.50& 0.08 & 0.75 & 0.12 & 1.11\\
    & 0.10 & 0.75 & 0.12 & 1.10\\
1.0 & 0.05 & 0.80 & 0.12 & 1.41\\
	& 0.10 & 0.78 & 0.12 & 1.42\\
2.0 & 0.05 & 0.84 & 0.12 & 2.23\\
    & 0.08 & 0.83 & 0.11 & 2.24\\
\end{tabular}
\end{ruledtabular}
%\end{table}

%\begin{table}[t]
\caption{\label{tab3}
The different-mesh results with $M_q=330$ MeV.
$\sigma_{\rm eff}$, $A_{\rm eff}$, and $C_{\rm eff}$ are the best-fit parameters in Eq.~(\ref{VQQq}) with the fit range of $R\le 1.2$ fm.
}
\begin{ruledtabular}
\begin{tabular}{ccccc}
Mesh& $\Delta z$ [fm]
& $\sigma_{\rm eff}$ [GeV/fm] & $A_{\rm eff}$ & $C_{\rm eff}$ [GeV]\\
\hline
64$\times$64& 0.12 & 0.72 & 0.12 & 1.07 \\
            & 0.15 & 0.72 & 0.12 & 1.08 \\
128$\times$128& 0.08 & 0.74 & 0.12 & 1.06\\
          	& 0.10 & 0.73 & 0.12 & 1.07\\
          	& 0.12 & 0.73 & 0.12 & 1.07\\
256$\times$256& 0.05 & 0.74 & 0.12 & 1.08 \\
\end{tabular}
\end{ruledtabular}
\end{table}

We fit the $QQq$ potential with 
\begin{eqnarray}
\label{VQQq}
V_{QQq}(R)=\sigma _{\rm eff}R-\frac{A_{\rm eff}}{R}+C_{\rm eff}
\end{eqnarray}
as an analogy of the $Q\bar{Q}$ potential.
The ``eff" means effective values including the light-quark effect.
The best-fit parameters are summarized in Table \ref{tab2}, and the obtained $QQq$ potential is shown in Fig.~\ref{fig3}. 
This fitting function is significantly suitable for $V_{QQq}$.
The effective Coulomb coefficient $A_{\rm eff}$ is almost the same value as $A_{3Q}$.
The effective string tension $\sigma _{\rm eff}$ is reduced about 10-20\% compared to the string tension $\sigma _{3Q}$ in the 3Q potential,
\begin{eqnarray}
\sigma _{\rm eff} < \sigma _{3Q} \simeq \sigma _{Q\bar{Q}} \simeq 0.89 \ {\rm GeV/fm}.
\label{SS}
\end{eqnarray}
This result means that the inter-two-quark confining force in baryons is reduced due to the light-quark existence.

To estimate the discretization error and the finite-volume effect, the results with different mesh numbers and mesh sizes are shown in Table \ref{tab3}.
These results are different only within a few percent.
We can see that our variational calculation is performed in enough large volume and enough fine mesh.

In our previous lattice QCD study \cite{YaUP}, we investigated the same $QQq$ potential in the region of $R\le 0.8$ fm.
For example, in the $M_q \simeq 1$ GeV case, the effective string tension is $0.75\pm 0.08$ GeV/fm.
In order to compare with the lattice QCD result, we calculate in the present potential model with the same condition, $M_q=1$ GeV and $R\le 0.8$ fm, and then find $\sigma _{\rm eff} \simeq 0.76$ GeV/fm.
Therefore, the calculation of our simple potential model almost reproduces the result of lattice QCD.

\subsection{The reduction of the effective string tension}

\begin{figure}[t]
\begin{center}
\includegraphics[scale=0.5]{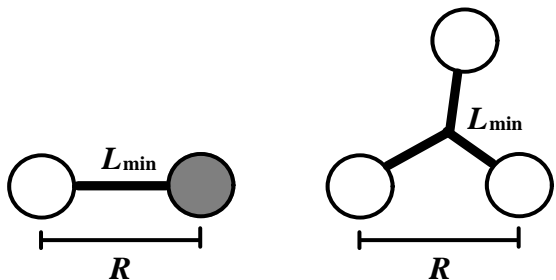}
\caption{\label{fig4}
The schematic figure of the flux-tube length $L_{\rm min}$ and the inter-quark distance $R$.
The two definitions are the same, $L_{\rm min} =R$, in the $Q\bar Q$ case (left).
These are different, $L_{\rm min} \neq R$, in the $3Q$ or $QQq$ case (right).
}
\end{center}
\end{figure}

To understand the reduction mechanism of $\sigma_{\rm eff}$ between the two heavy quarks, we compare the definition of the string tension with that of the effective string tension.
The string tension is the proportionality coefficient of the color flux-tube length $L_{\rm min}$ in the confinement potential.
Compared with this, the effective string tension is defined as the effective confinement force between two quarks in baryons in terms of the inter-two-quark distance $R$ .
The schematic figure is depicted in Fig.~\ref{fig4}.
In the $Q\bar{Q}$ case, $L_{\rm min} =R$ and the two definitions lead no difference, but in the $3Q$ or $QQq$ case, these are different from each other, $L_{\rm min} \neq R$.
The $QQq$ flux-tube length is related to $R$ through complicated light-quark dynamics, and then the $QQq$ effective confinement potential is not necessarily a linear function of $R$.
Our results show that the effective confinement potential is almost proportional to $R$, at least in the region of $R\le 1.2$ fm, with the proportionality coefficient $\sigma_{\rm eff}$, which is reduced compared to the string tension of mesons and the static cases.
We conjecture that the reduction of $\sigma_{\rm eff}$ originates from this difference between the inter-heavy-quark distance $R$ and the flux-tube length $L_{\rm min}$.

\begin{figure}[t]
\begin{center}
\includegraphics[scale=1]{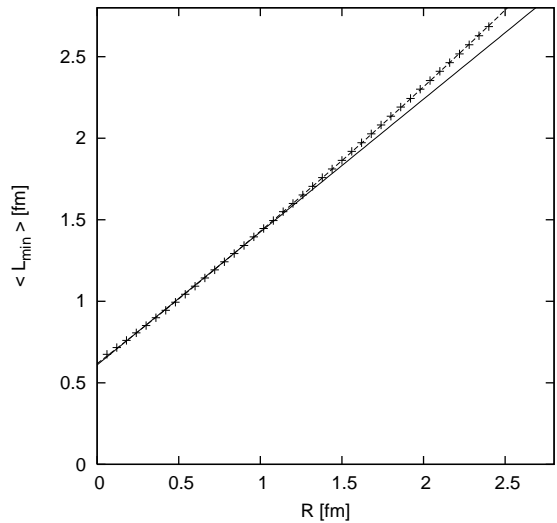}
\caption{\label{fig5}
The relation between $\langle L_{\rm min}\rangle$ and $R$ with $M_q=330$ MeV.
The symbol + denotes the result from the variational calculation. 
The solid line is the best-fit function of $b_0+b_1R$ with the fit range of $R\le 1.2$ fm, and the dashed line is that of $b_0+b_1R+b_2R^2$ with $R\le 2.4$ fm.
}
\end{center}
\end{figure}

The relation between $R$ and the $QQq$ flux-tube length can be easily calculated in our potential model.
The expectation value of the $QQq$ flux-tube length $\langle L_{\rm min} \rangle$ is shown in Table \ref{tab1} and Fig.~\ref{fig5}.
As shown in Fig.~\ref{fig5} (solid line), $\langle L_{\rm min} \rangle$ is well described as $\langle L_{\rm min} \rangle \simeq b_0+b_1R$ in the range of $R\le 1.2$ fm, and its best-fit parameters are $b_0\simeq 0.61$ fm and $b_1\simeq 0.81$.
Apart from the light-quark kinetic energy and the Coulomb contribution, we can explain the reduction of $\sigma_{\rm eff}$ by the functional form of $\langle L_{\rm min} \rangle$, that is,
\begin{eqnarray}
\langle \sigma _{3Q}L_{\rm min} \rangle &\simeq& (b_1\sigma _{3Q})R+b_0\sigma _{3Q} \nonumber\\
&\simeq& \sigma_{\rm eff}R +{\rm const.}
\label{S3QLmin}
\end{eqnarray}
Thus $\sigma_{\rm eff}$ is reduced as $\sigma_{\rm eff}\simeq b_1\sigma _{3Q}$, and $b_1(<1)$ means the reduction rate of $\sigma_{\rm eff}$.

From the different $M_q$ results, we see the $M_q$-dependence of the system.
Intuitively, in the $M_q\to \infty$ limit, the $QQq$ system becomes the static $3Q$ system and then $\sigma _{\rm eff}$ approaches to $\sigma _{3Q}$.
It is expected that, when $M_q$ increases, the reduction is weakened and $\sigma _{\rm eff}$ increases.
Similarly, since quarks are not spatially extended at all in the $3Q$ case, the light-quark spatial distribution becomes compact for large $M_q$.
We can check these behaviors in Table \ref{tab1} and \ref{tab2}.

Now we investigate the contribution of the Coulomb terms in Eq.~(\ref{V123}).
We artificially set $A_{3Q}=0$, and calculate in the same scheme as above.
We fit this ``No-Coulomb" $QQq$ potential with a functional form
\begin{eqnarray}
V_{QQq}^{\rm NC}(R)=\sigma _{\rm eff}^{\rm NC}R+C_{\rm eff}^{\rm NC}.
\label{VNC}
\end{eqnarray}
The best-fit parameters are shown in Table \ref{tab4}.
Also in this case, the effective string tension $\sigma _{\rm eff}^{\rm NC}$ is reduced, and roughly equals to $\sigma _{\rm eff}$. 
Therefore, the essential reason for the reduction of $\sigma _{\rm eff}$ is the geometrical difference between $L_{\rm min}$ and $R$ rather than the Coulomb contribution.

\begin{table}[t]
\caption{\label{tab4}
The ``No-Coulomb" ($A_{3Q}=0$) $QQq$ potential results with the 128$\times$128-mesh.
$\sigma_{\rm eff}^{\rm NC}$ and $C_{\rm eff}^{\rm NC}$ are the best-fit parameters in Eq.~(\ref{VNC}) with the fit range $R\le 1.2$ fm.
}
\begin{ruledtabular}
\begin{tabular}{cccc}
$M_q$ [GeV] & $\Delta z$ [fm] & $\sigma_{\rm eff}^{\rm NC}$ [GeV/fm] & $C_{\rm eff}^{\rm NC}$ [GeV]\\
\hline
0.33& 0.10 & 0.69 & 1.19\\
	& 0.12 & 0.69 & 1.19\\
0.50& 0.08 & 0.70 & 1.25\\
    & 0.10 & 0.70 & 1.25\\
\end{tabular}
\end{ruledtabular}
\end{table}

\begin{figure}[t]
\begin{center}
\includegraphics[scale=1.2]{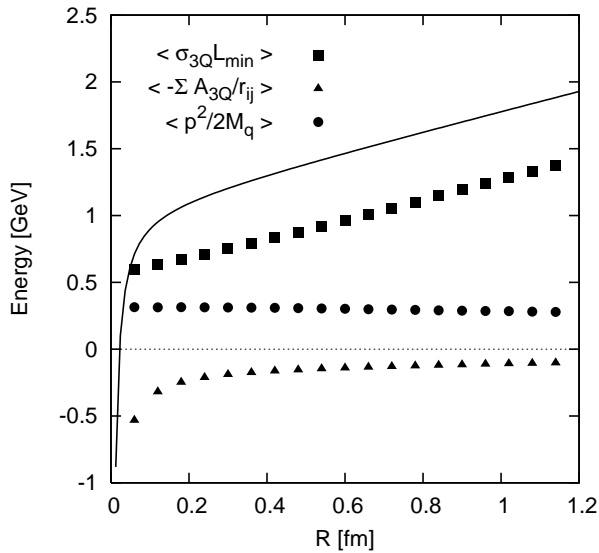}
\caption{\label{fig6}
The component of the $QQq$ potential with $M_q=330$ MeV.
The solid curve is the best-fit function $V_{QQq}(R)$ of Eq.~(\ref{VQQq}), and it is given as $V_{QQq}(R)=M_q+\langle \vec{p}^2/2M_q+\sigma_{3Q}L_{\rm min}-\sum_{i<j} A_{3Q}/r_{ij}\rangle$.
}
\end{center}
\end{figure}

In Fig.~\ref{fig6}, we separate the $QQq$ potential with $M_q=330$ MeV into the contributions from the expectation value of the light-quark kinetic term, the $3Q$ confinement potential term, and the Coulomb term.
The resulting $QQq$ potential is given as $V_{QQq}(R)=M_q+ \langle -\frac{1}{2M_q}\frac{\partial^2}{\partial \vec{r}_3^2}+\sigma_{3Q}L_{\rm min}-\sum_{i<j} \frac{A_{3Q}}{r_{ij}}\rangle$.
We can confirm that the confinement part of the $QQq$ potential originates mainly from the $3Q$ confinement potential contribution.

\subsection{The detail of the effective string tension}

We note that the $QQq$ potential is not perfectly fitted with Eq.~(\ref{VQQq}).
For example, in the large $R$ limit, $R$ will approach to the flux-tube length $\langle L_{\rm min}\rangle$, so that $\sigma _{\rm eff}$ increases and approaches to $\sigma _{3Q}$.
Thus $\sigma _{\rm eff}$ should depend on $R$.
To investigate the more accurate $R$-dependence, we enlarge the region of $R$ to 2.4 fm.
As mentioned above, $\langle L_{\rm min} \rangle$ is determined by a nontrivial light-quark dynamics.
Then, in general, it is a function with higher-power terms of $R$,
\begin{eqnarray}
\langle L_{\rm min} \rangle=b_0+b_1R+b_2R^2+b_3R^3+ \cdots .
\end{eqnarray}
If we fit $\langle L_{\rm min} \rangle$ with up to the quadratic term with $M_q=330$ MeV and $R\le2.4$ fm, we find that the best-fit parameters are $b_0\simeq 0.62$ fm, $b_1\simeq 0.77$, and $b_2\simeq 0.037$ fm$^{-1}$.
Its best-fit function is shown in Fig.~\ref{fig5} (dashed line).
In this region, the linear function deviates from the result, and this quadratic function seems suitable for fitting.
In this case, if the light-quark kinetic energy and the Coulomb contribution is ignored, the effective string tension is roughly written as
\begin{eqnarray}
\sigma _{\rm eff} (R) \equiv \frac{\partial V_{QQq} (R)}{\partial R}\Big |_{\rm IR} \simeq b_1\sigma_{3Q}+2b_2\sigma_{3Q}R, 
\label{Seff}
\end{eqnarray}
where ``IR" means the infrared region.
Thus the effective string tension actually depends on $R$.
To investigate the $R$-dependent effective string tension with the whole interaction of Eq.~(\ref{V123}), we fit the $QQq$ potential with 
\begin{eqnarray}
\label{VQQq2}
V_{QQq} (R)=(b_1+b_2R)\sigma_{3Q}R-\frac{A_{\rm eff}}{R}+C_{\rm eff}
\end{eqnarray}
in the range of $R\le2.4$ fm.
The results are summarized in Table \ref{tab5}.
From the fact that $b_2$ is positive, $\sigma _{\rm eff}(R)$ is an increasing function of $R$, as is expected.
Of course, this argument is restricted in the present range of $R$.
For larger $R$, higher-power terms are to be included.

\begin{table}[t]
\caption{\label{tab5}
The best-fit parameters of Eq.~(\ref{VQQq2}) with the fit range of $R\le 2.4$ fm.
The calculation is done with $M_q=330$ MeV and the 128$\times$128-mesh.
}
\begin{ruledtabular}
\begin{tabular}{cccccc}
$\Delta z$ [fm]& $b_1$ & $b_2$ [fm$^{-1}$] & $A_{\rm eff}$ & $C_{\rm eff}$ [GeV]\\
\hline
0.10 & 0.78 & 0.031 & 0.12 & 1.08 \\
0.12 & 0.78 & 0.033 & 0.12 & 1.08 \\
\end{tabular}
\end{ruledtabular}
\end{table}

\subsection{Comments on the relativistic corrections}

The non-relativistic quark model has achieved considerable success in explaining low-energy hadron properties.
However, the relativistic treatment for the light quark can give rise to the quantitative correction to some properties.
We give a brief comment on the relativistic correction to our result.

In the relativistic formalism for the light quark, the Hamiltonian (\ref{Hami}) is modified as
\begin{eqnarray}
\label{relHami}
H&=&\sqrt{M_q^2+\vec{p}^2}+V(\vec{r}_1,\vec{r}_2,\vec{r}_3) ,
\end{eqnarray}
where $\vec{p}$ is the light-quark momentum.
The interaction term $V(\vec{r}_1,\vec{r}_2,\vec{r}_3)$ is again the general three-quark interaction in $QQq$ systems.

One of the relativistic corrections is the higher-order kinetic term from $(M_q^2+\vec{p}^2)^{1/2}-(M_q+\vec{p}^2/2M_q)$.
If the higher-order terms do not change the $R$-dependence of the $QQq$ potential, the non-relativistic formalism works well.
As shown in Fig.~\ref{fig6}, the leading-order kinetic term $\vec{p}^2/2M_q$ is almost $R$-independent, compared to the typical $R$-dependence of the energy, namely $\sigma_{3Q}\simeq0.89$ GeV/fm.
We expect that the higher-order kinetic terms also give rise to small $R$-dependent contributions, at least in the large $M_q$ case. 

Another possible relativistic correction is modification to $V(\vec{r}_1,\vec{r}_2,\vec{r}_3)$.
While we adopt the static $3Q$ potential as the three-quark interaction in $QQq$ systems, it can be changed by the light-quark relativistic effect.
Several works have dealt the inter-quark potential in mesons and baryons with finite-mass quarks \cite{Br05,Br052}.
The finite-mass-quark effect on the potential is rather complicated, and nontrivial $M_q$-dependence can occur.

For the precise statement and quantitative accuracy, the relativistic treatment is desired, particularly for the realistic light-quark mass case, i.e., $M_q\simeq 330$ MeV.

\section{Summary and Conclusion}

From the above considerations, we can conclude that, when finite-mass quarks exist in baryons, the effective string tension between the other two quarks is reduced compared to the static case. 
The essential reason for the reduction is the {\it geometrical difference} between the inter-quark distance $R$ and the flux-tube length $\langle L_{\rm min} \rangle$.
As can be seen from Fig.~\ref{fig4}, this geometrical difference always exists for more than two-quark system, regardless of whether the heavy-quark mass is infinite or finite.
Then this is expected to hold not only for $QQq$ systems but also for ordinary baryons, such as nucleons.
Of course, the quantitative difference would exist between $QQq$ systems and ordinary baryons.
In addition, also in multi-quark systems, the effective string tension can be changed due to the existence of other light quarks.
In such systems, the effective string tension between two finite-mass quarks would include more complicated finite-mass effects.

In summary, we have studied the $QQq$ potential with the non-relativistic potential model, and have investigated the finite-mass-quark motional effect on the inter-heavy-quark potential in baryons.
We have found the significant reduction of the effective string tension $\sigma _{\rm eff}$ between the two heavy quarks in $QQq$ systems, compared to string tension $\sigma _{3Q}$ in the static $3Q$ case.
The effective string tension $\sigma _{\rm eff}$ depends on the light-quark mass $M_q$, and slightly depends on $R$.
The finite-mass-quark existence reduces the effective confinement force, which is conjectured to be a general property for baryons.

\section*{ACKNOWLEDGEMENTS}

%\begin{acknowledgements}
H.~S.~was supported in part by a Grant for Scientific Research [(C) No.19540287] from the Ministry of Education, Culture, Sports, Science and Technology of Japan.
%\end{acknowledgements}

\end{document}